# Resolving transitions in the mesoscale domain configuration in VO$_2$ using laser speckle pattern analysis

Katyayani Seal[1,2], Amos Sharoni[3,4], Jamie M. Messman[5], Bradley S. Lokitz[5], Robert W. Shaw[6], Ivan K. Schuller[4], Paul C. Snijders[1,2], Thomas Z. Ward[1]

[1] Materials Science and Technology Division, Oak Ridge National Laboratory, Oak Ridge, TN 37831, USA
[2] Department of Physics & Astronomy, University of Tennessee, Knoxville, TN 37996, USA
[3] Department of Physics and Institute of Nanotechnology, Bar Ilan University, Ramat Gan 52900, Israel
[4] Department of Physics and Center for Advanced Nanoscience, University of California San Diego, La Jolla, California 92093, USA
[5] Center for Nanophase Materials Sciences, Oak Ridge National Laboratory, Oak Ridge, TN 37831, USA
[6] Chemical Sciences Division, Oak Ridge National Laboratory, Oak Ridge, Tennessee 37831, USA

**The configuration and evolution of coexisting mesoscopic domains with contrasting material properties are critical in creating novel functionality through emergent physical properties. However, current approaches that map the domain structure involve either spatially resolved but protracted scanning probe experiments without real time information on the domain evolution, or time resolved spectroscopic experiments lacking domain-scale spatial resolution. We demonstrate an elegant experimental technique that bridges these local and global methods, giving access to mesoscale information on domain formation and evolution at time scales orders of magnitude faster than current spatially resolved approaches. Our straightforward analysis of laser speckle patterns across the first order phase transition of VO$_2$ can be generalized to other systems with large scale phase separation and has potential as a powerful method with both spatial and temporal resolution to study phase separation in complex materials.**

Self-organization of mesoscopic domains that arise from underlying microscopic interactions is ubiquitous in condensed matter physics, leading to emergent macroscopic properties in materials ranging from elementary ferromagnetic materials[1] to electronic phase separation in complex oxides[2,3,4] Specifically, the many fascinating macroscopic properties found in transition metal oxides are often intimately related to their domain structure, such as colossal magnetoresistance in phase separated manganites[2] and stripe order in high temperature superconducting cuprates[3,4]. Beyond their fundamental importance, the presence of coexisting domains offers the opportunity to manipulate macroscopic materials properties and engineer new functionality through perturbations that affect the relative stability of the different domains, such as demonstrated in the proposed racetrack memory in ferromagnetic materials[5] and multibit information processing in phase separated oxides[6].

While these materials consist of domains in their ground state, a much broader class of materials, those exhibiting first order phase transitions, feature coexisting domains in a temperature range near their metal-insulator transition temperatures, $T_{MIT}$. A current example enjoying intense interest is VO$_2$, which has a first order metal-insulator transition at ~340 K. In one proposal, facilitated by the correspondence of optical wavelengths with the domain sizes of the coexisting metallic and insulating phases in VO$_2$ near $T_{MIT}$, these domains have been highlighted for use in a tunable optical metamaterial with applications ranging from perfect absorbers to chemical sensing[7,8]. In a temperature range near $T_{MIT}$, VO$_2$ is closely balanced between a low temperature insulating state and a high temperature metallic state; in this region, even small perturbations can have a tremendous influence on macroscopic properties by slightly modifying the domain configuration. It is therefore critically important to understand the domain ordering process to take advantage of the intrinsic tunability offered by these materials. This necessitates an experimental approach with a spatial resolution on the order of the domain size or smaller, and a measurement speed significantly faster than the domain evolution near the transition. The techniques commonly employed to this end rely on scanned probe experiments, such as scanning near-field infrared microscopy[9], magnetic force microscopy (MFM) and electron microscopy[10,11]. These scanned probe techniques are inherently time consuming and extremely local, making it difficult to map domain evolution accurately, especially when varying the temperature. Conventional optical techniques such as UV-Vis absorption and ellipsometry [9,12,13] can provide vital information on the far field (averaged) effective dielectric properties at very short time scales, but they are spatially averaged global measurements that do not provide the spatial resolution necessary for an understanding of domain configurations.

We introduce an elegant approach that bridges these local and global techniques and gives access to mesoscale information on domain formation and evolution. Due to the large dielectric contrast between the insulating and metallic domains in $VO_2$, the optical response of $VO_2$ across the MIT is manifested directly in its laser speckle pattern measured using wavelengths in the visible regime. A statistical analysis of the laser speckle intensity distribution allows us to map the MIT with high accuracy in the temperature domain during continuously changing temperature while retaining spatial information on the coexisting phases. The statistical analysis can be performed in real time, providing immediate information on the dielectric contrast of the metallic and insulating phases on the $VO_2$ surface. Our approach can be generalized to other systems with large scale phase separation and has the potential of being a powerful method of analyzing phase separated materials using their optical response.

**Experiment**

In the experiment, a 100 nm $VO_2$ thin film was deposited by reactive RF-sputtering on a r-cut sapphire substrate, resulting in a single phase epitaxial layer; growth details presented elsewhere[14]. The sample was mounted on a heating stage. Linearly polarized laser light of intensity $< 0.5$ mW/cm$^2$ at three different wavelengths (488 nm, 633 nm and 800 nm) was collinearly incident on the sample with a spot size of 1.2 mm at 45 degrees to the sample plane and the reflected light was imaged directly onto a cooled CCD detector (Princeton Instruments NTE CCD 512 EBFT GR1) located at a distance of 50 cm at a slightly oblique angle (3 degrees off the specular direction) sufficient to exclude the principal reflected specular maximum and record only the diffuse speckle portion of the reflected light. The pixel count was 512 x 512 pixels over a 0.5 cm x 0.5 cm illumination area of the CCD, which resulted in approximately 9 pixels per speckle for the shorter wavelengths. The speckle patterns were recorded with increasing temperature (~1 $^O$C/sec) approximately every 0.3 $^O$C while simultaneously measuring the in-plane resistance outside of the laser spot to trace the metal-insulator transition. The CCD collection time for a single speckle pattern was 1 msec. The speckle patterns were then statistically analyzed for their intensity distributions. The optical constants of the $VO_2$ film and their variation with temperature and wavelength were measured using a J.A. Woollam M-2000U variable angle spectroscopic ellipsometer over a wavelength range of 245-999 nm with a Linkam scientific heating stage (THMS600).

**Results and Discussion**

First we performed spectroscopic ellipsometry measurements as a function of temperature to optically characterize the metal-insulator transition (MIT) of the sample. The ellipsometry data were fitted with a Lorentz oscillator model to obtain $\varepsilon_1(\lambda)$ and $\varepsilon_2(\lambda)$ as shown in Fig 1(a-b). From these data we obtained $\varepsilon(T)$ at each wavelength, shown in Fig 1(c-d). Here, a regression analysis was used to simultaneously analyze data taken at three wavelengths and the mean square error (MSE) was used to quantify the difference between the experimental data and model. The unknown parameters were allowed to vary until the minimum MSE was reached. It is clear that the MIT starts close to 40 °C. The actual sample temperature lags behind the recorded temperature in the ellipsometry data since the thermocouple was placed on the heating stage and not directly on the sample. The changes in $\varepsilon_1(T)$ and $\varepsilon_2(T)$ for the three laser wavelengths reveal that the dielectric constant transitions between two dominant values starting at 40 °C and completing at approximately 55 °C. At 800 nm, the value of $\varepsilon_1$ decreases strongly at the transition but remains above zero, indicating weak metallic behavior. This is expected since this illumination is still above the VO$_2$ plasmon frequency[7]. This implies that for all our subsequent speckle measurements (see below), the surface scattering is not affected by plasmon absorption. At around 1000 nm, $\varepsilon_1$ becomes negative above the MIT, demonstrating the anticipated metallic behavior of VO$_2$ at high temperatures. For wavelengths between 540 and 650 nm, $\varepsilon_1$ actually increases with temperature across the MIT. The value of $\varepsilon_2$ consistently drops across the MIT at all wavelengths. At the temperature where the fraction of metal domains equals that of insulating domains, i.e. near the MIT,[15] we obtain the effective dielectric constant using the Maxwell Garnet effective medium approximation: 488 nm: 0.2+3i, 633 nm: 0.7+3.6i, 800 nm: 0.02+2.2i.

Optical speckle patterns are a result of the interaction of laser light with the local optical features of a surface in terms of the spatial distribution of dielectric contrast. The speckle patterns from the VO$_2$ surface show a variation of the average brightness, speckle position and image contrast as a function of temperature and wavelength. A simplified diagram of the experimental setup and an example of a typical speckle pattern is given in Fig. 2a. To quantify speckle pattern changes, the intensity statistics were calculated from each image and plotted as a function of temperature in Fig. 2(b-g). The statistical properties of the speckle show a transitional behavior close to the $T_{MIT}$ evident in the resistance data. The mean value of the speckle intensity MI = $\langle I \rangle$ and the variance V = $(\langle I^2 \rangle - \langle I \rangle^2)$ show the onset of the transition in the form of a significant change in magnitude within a temperature range of 57-60 $^O$C. As in the temperature dependent resistance, the first order phase transition produces a hysteresis in the MI and the variance although with a narrower width. The decrease in the MI is easily understood since the overall reflectivity is expected to increase across the transition due to the growth of the metallic phase volume which has higher reflectivity. This implies an increase in the principal reflected specular intensity and a corresponding decrease in the measured diffused speckle intensity. The transition in the optical measurements occurs at a higher temperature than the electronic transition on both heating and cooling. This can be attributed to the fact that in percolative systems, low resistance pathways can be present already at low metal domain coverages of 20%[16]. The optical response, where the effective lengthscale of the metallic domains needs to be on the order of the wavelength used[15], requires a higher metal domain coverage for it to be sensitive to the domain configuration.

The variance is a measure of the amount of fluctuation in the speckle pattern, which corresponds to the amount of dielectric contrast in the system. In our experiment the film thickness is much smaller than the wavelength of the incident light. Hence the surface topography is smooth on the order of the wavelengths used and the changes in dielectric contrast are dominated by the contrast between lateral metallic and insulating phase domains, and not by physical surface roughness. Both far above and far below the MIT, the sample consists of a single phase and no phase domain induced dielectric scattering should occur. The enhanced variance near the transition as seen in the 633 nm data and weakly in the 800 nm data demonstrates the presence of a mixed phase domain configuration near $T_{MIT}$ with length scales on the order of $\lambda/2$ [17]. The decrease in the absolute value of the variance in the metallic phase far above the MIT as compared to that in the insulating phase far below the MIT, despite the single phase nature of the sample at these temperatures, suggests that the increased specular scattering of the incident light in the metallic phase results in a lower speckle intensity with an associated lower contrast.

Importantly, the speckle pattern allows us to glean information about the configuration of metallic and insulating domains in the sample. This is clear from a number of observations. First, across the MIT, the variance decreases by 2%, 4%, and 15% under illuminations of 488 nm, 633 nm, and 800 nm, respectively. This increase is produced by the higher sensitivity of our experiment to surface dielectric features having length scales on the order of the wavelength used. The larger change in variance for longer wavelengths therefore indicates that the temperature dependent domain configurations on the $VO_2$ surface are dominated by correlation lengths which are closer in magnitude to the longer laser wavelengths. Second, the increased width of the transition region at 488 nm as compared to the longer wavelengths (Fig. 2) indicates that the smaller correlation lengths in the insulating and metallic domains making up the dielectric landscape exist in a larger temperature range around the transition, while larger sized domains are dominant in the middle of the transition, consistent with domain nucleation and growth[9,14]. Note that this persistence to higher temperatures also confirms that the low photon flux used in this experiment does not cause a photon-induced decrease in transition temperature as it would then be expected that the transition would be completed at progressively lower temperatures for decreasing wavelengths. Third, a comparison of the temperature dependent MI of the speckle pattern intensity in Fig 3, demonstrates that the onset of the transition in the MI increases whereas its transition width decreases with increasing wavelength.

Laser speckle analysis has been used in the past as a surface roughness characterization technique[23-25], and it was found that different scattering processes give rise to distinct speckle correlation lengths. Multiple scattering processes can be differentiated from single or weak scattering processes through the intensity or field statistics[18,19]. The first order intensity statistics of the electromagnetic field scattered from randomly rough conducting surfaces are a function of the surface root mean square (rms) roughness[18] and we will use this next to extract a measure of the dielectric contrast from the data. In our experiment we use a large oblique angle of incidence, and measure the speckles in an area near the specularly reflected spot. Hence we are mainly sensitive to weak scattering[17,20], and one can assume that the wavelength is much smaller than the scattering mean free path. In this weak scattering limit, it is sufficient to measure the integrated scattered intensity, $P_{scatt}$, which is directly proportional to the mean-square roughness as follows: $P_{scatt}/P_{tot} = (4\pi\sigma \cos\theta_i \lambda)^2$, where $P_{tot}$ is the total reflected intensity, and $\sigma$, $\theta_i$ and $\lambda$ are the rms roughness, angle of incidence, and wavelength of the incident light, respectively. The rms roughness obtained in this way is related to the surface correlation length[18], which in our experiment is dominated by the dielectric contrast of the $VO_2$ sample. While we measure only a fraction of the non-specularly scattered intensity, the number of speckles imaged in our experiment is sufficiently high to use the statistics obtained from the integrated fractional scattering intensity, $P_{f,scatt}$, to provide a measure of the evolution of the dielectric contrast with temperature. Using the above expressions, we calculate the temperature dependence of $\sigma_\Delta = \sigma_{f,T_n}/\sigma_{f,T_{n-1}}$ as a normalized measure for the dielectric contrast for the three wavelengths, see Fig. 3(c), where $\sigma_{f,T_n}$ is the roughness at the $n^{th}$ value of the temperature in our measurement series, $T_n$, obtained from the fractional scattering intensity. $\sigma_\Delta$ peaks at the MIT and is greater for increasing wavelengths. For shorter wavelengths, the optical response is more sensitive to domains of smaller (order of wavelength) size. Therefore the increased magnitude of $\sigma_\Delta$ at longer wavelengths indicates that at the MIT the combined effect of the dielectric contrast between metallic and insulating domain, as well as their average spatial extent, are both maximized. This is consistent with a percolation model of domain growth towards a first order metal-insulator transition[15,21,22,14]. Although the shorter wavelength measurements are dominated by noise, we do observe a shallow peak at 633 nm, suggesting that the changes in effective dielectric contrast are smeared out over a larger temperature range as also observed with the variance data. The lower $\sigma_\Delta$ also reveals that dielectric features of the order of the shorter wavelengths have a smaller impact on the optical response and therefore likely exist in lower concentrations assuming their dielectric metal-insulator contrast is the same.

From the sharp transition in the speckle statistics near the transition temperature, it is clear that the presence of nucleated metallic domains causes a sharp change in the optical response. Our diffraction limited speckle measurement is clearly sensitive enough to differentiate the dielectric features on the $VO_2$ surface. Figures 3(a) and (b) show that shorter wavelengths result in a measured transition width that is broader and has an earlier onset of transition for both the MI and the variance of the speckle pattern. The observed wavelength dependence of the variance thus shows that with the wavelengths used in our experiments we are sensitive to the length scales of the phase separated domains in $VO_2$ near its MIT.

These wavelength dependent speckle pattern snapshots therefore allow one to study the nucleation and growth of dielectrically contrasting domains as a function of temperature. Below we discuss an expanded analysis using second order intensity statistics.

While first order intensity statistics allow one to extract the roughness of the dielectric landscape, second order intensity statistics such as autocorrelation analysis is very sensitive to small variations in speckle sizes. Speckle correlation function coefficients depend on the distribution of the dielectric constant at the surface, with the spatial resolution for roughness features being constrained by the diffraction limit[17,23,24,25]. The autocorrelation function thus provides an independent spatial analysis of the dominant length scales present in each speckle pattern which are related to the morphology of the dielectric landscape on the VO$_2$ surface. We calculate the 2D autocorrelation plot from each image as well as the cross correlation of consecutive pairs of image recorded with increasing temperature. To effectively analyze this, first the 2D autocorrelation, $C_{T_n}$, from each image (at each temperature) was obtained using $C_{T_n}(dx, dy) = \langle \delta I(x,y) \delta I(x+dx, y+dy) \rangle$, where $\delta I(x,y) = (I(x,y) - \langle I(x,y) \rangle)/\langle I(x,y) \rangle$. Next, the $x$ and $y$ cross sections $C_{T_n}(dx, 0)$ or $C_{T_n}(0, dy)$ of each image were obtained and subtracted from their consecutive counterpart recorded at the next temperature, $C_{sub}(dx) = C_{T_n}(dx, 0) - C_{T_{n-1}}(dx, 0)$. The maximum value of the shift in the $x$ and $y$ directions ($dx$ and $dy$, respectively) were about half the image size (~ 250 pixels). The results for wavelengths 633 nm and 800 nm are plotted in Fig 4(a)-(d). The subtracted autocorrelation cross sections, $C_{sub}$, are essentially flat at most temperatures. This is especially true for the area of the central maximum at all temperatures, which confirms that the average speckle size stays constant with temperature at each wavelength. Near the MIT transition temperature, a secondary peak develops away from the central maximum region, as seen most prominently in Fig 4(a) and (b) at T = 57 °C for an illumination wavelength 633 nm and at a slightly higher temperature of 59 °C for 800 nm. The secondary maxima are a signature of long range correlation, which indicates a decreased scattering mean free path[25] and increased scattering strength. Additional small maxima at different positions are also noticeable at temperatures 61 °C for 800 nm and 62 °C for 633 nm. We attribute these maxima to changes in the metallic domain size and the resulting effective dielectric contrast near the transition. Coupled with the peak in effective dielectric roughness variation observed in Fig. 3(c), this indicates that the dielectric contrast is maximum at the MIT as expected from the percolation model. At shorter wavelengths the maxima occur across a wider temperature range. This is due to the fact that detected domain sizes are smaller and will therefore be detected further away from the transition point, similar to the larger width in the variance for shorter wavelengths (see Fig. 2).

**Conclusion**

We have investigated the surface dielectric contrast of self-organized mesoscopic domain configurations in VO$_2$ across the MIT using a simple scheme involving laser speckle intensity statistics in the weak scattering regime. Our experiment shows that the wavelengths used in our analysis are sensitive to the dielectric surface contrast of the inherent metallic and insulating domains and their spatial extent (lengthscales) in the phase separated state near the T$_{MIT}$ of VO$_2$. The speckle statistics, both first and second order, accurately reveal the MIT and exhibit hysteretic behavior. The maximum of the surface dielectric contrast and scattering strength at the transition points corroborate this. The ellipsometry data shows a strong dual-phase nature and a large change in the dielectric constant across the transition. Our work on VO$_2$ demonstrates that in large scale phase separated materials with sufficient dielectric contrast, laser speckle experiments offer a bridge between slow but high resolution scanned probe methods, and fast but spatially averaged optical characterization. An analysis of speckle statistics can be performed in real time, and offers a straightforward method to investigate multi-domain dynamics at time scales orders of magnitude faster than state-of-the-art scanning probe techniques.

**Acknowledgments**

Research supported by the US Department of Energy (DOE), Basic Energy Sciences (BES), Materials Sciences and Engineering Division, (PCS, TZW) and Chemical Sciences, Geosciences, and Biosciences Division (RWS). Ellipsometry measurements (JMM, BSL) were conducted at the Center for Nanophase Materials Sciences, which is sponsored at Oak Ridge National Laboratory by the Scientific User Facilities Division, Office of BES, US DOE. Partial support was also given by LDRD Program at ORNL (KS). Partial support was also given by, U.S. Department of Energy, BES-DMS funded by the Department of Energy's Office of Basic Energy Science, under grant DE FG03 87ER-45332 (AS, IKS).

**Author contributions**

Sample growth was done by AS and IKS. Ellipsometry was done by KS, JMM and BSL. Speckle pattern collection was done by KS, RWS and TZW. Speckle pattern analysis was done by KS. KS, PCS and TZW designed the experiments and wrote the paper.

**Additional information**

Reprints and permissions information is available online at www.nature.com/reprints.

Correspondence and requests for materials should be addressed to PCS (snijderspc@ornl.gov) and TZW (wardtz@ornl.gov).

**Competing financial interests**

The authors declare no competing financial interests.

**Figures**

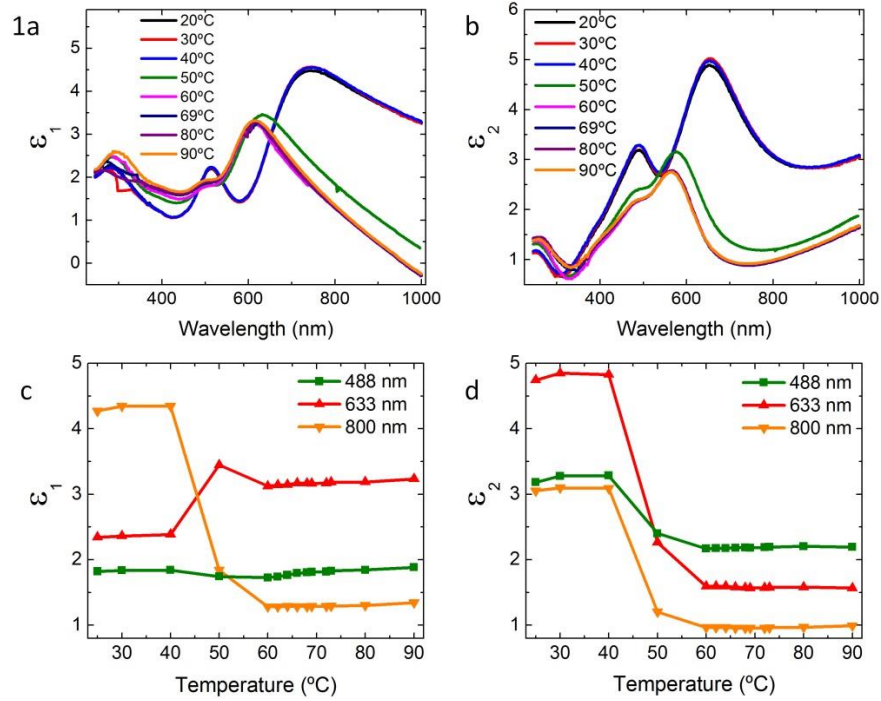

Figure 1. Ellipsometry data above, at, below the $T_{MIT}$ for $VO_2$. The (a) real and (b) imaginary parts show a variation in the dielectric permittivity with wavelength at different temperatures. The (c) real and (d) imaginary parts of the dielectric permittivity and their variation with temperature at the illumination wavelengths used in the speckle measurements show dual-phase behavior across the transition temperature.

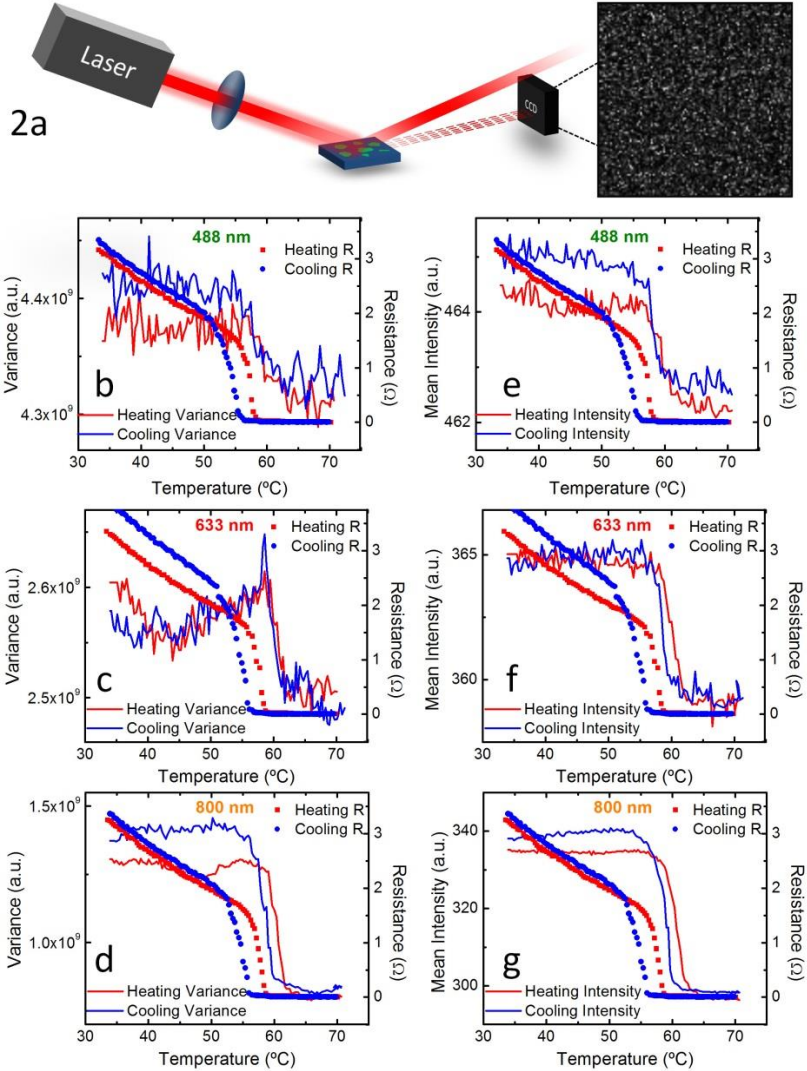

Figure 2. (a) Simplified diagram of experimental setup for speckle pattern collection. Polarized laser light is focused onto a sample using a lens and is reflected from the $VO_2$ surface where green and blue represent regions of different dielectric values. The reflected diffuse (dashed) speckle pattern is sampled using a CCD, the specularly reflected light (solid) is not collected. The image gives an example of a typical speckle pattern under 800 nm illumination at 60 °C. (b)-(g) Variance (b, c, d) and mean intensity (e, f, g) as a function of temperature plotted for three illumination wavelengths, 488 nm, 633 nm and 800 nm. Also plotted is the d.c. resistance. The shaded regions indicate the width of the transition in the optical data of each respective panel. The relatively increased noise in the variance for the shorter wavelengths is due to the smaller number of pixels per speckle at these shorter wavelengths.

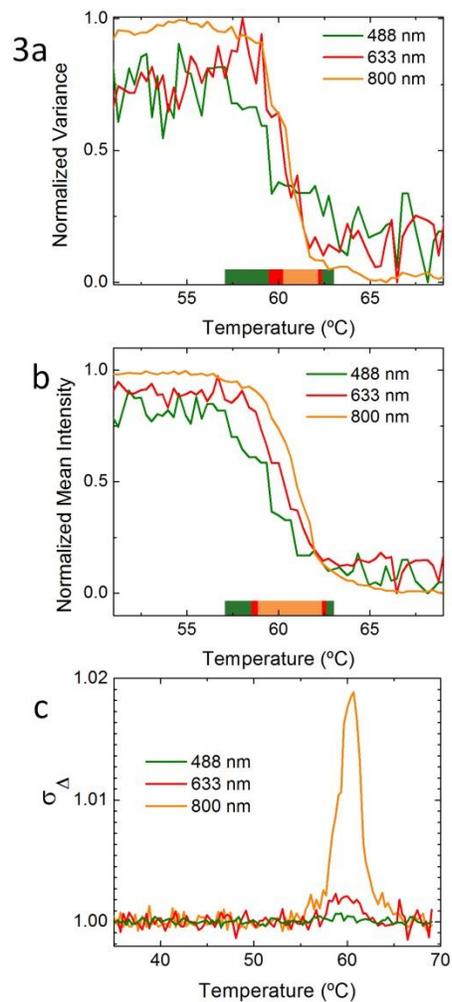

Figure 3. (a) Normalized variance and (b) normalized mean value of intensity plotted for three illumination wavelengths, 488 nm, 633 nm and 800 nm for increasing temperature. Plots focus on region near the $T_{MIT}$ showing the variation of the transition point. The normalization was implemented by setting the maximum and minimum value of the variance or mean intensity observed in the full temperature range to 1 and 0, respectively. Color bars indicate temperature range of observed transition for each wavelength. (c) Surface roughness ratio $\sigma_\Delta$ for the three wavelengths showing a peak at the transition point.

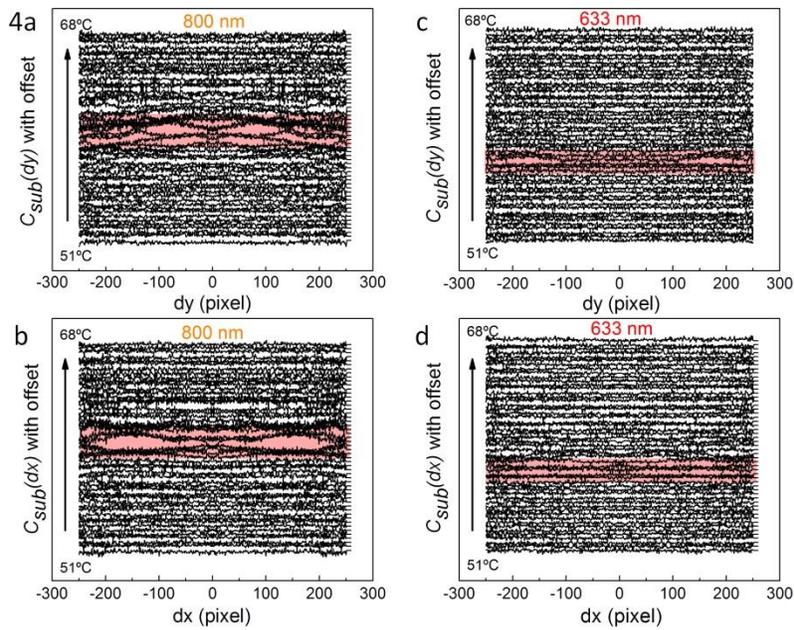

Figure 4. Subtracted autocorrelation functions for different temperatures under 800 nm illumination along the *x* (a) and *y* (b) axes with red highlighted regions showing transition between 58.1 °C and 62.2 °C. The shifts in the *x* and *y* axes, *dx* and *dy*, are in pixels. Subtracted autocorrelation functions across the same temperature range under 633 nm illumination along *x* (c) and *y* (d) with red highlighted regions showing transition between 56.8 °C and 58.9 °C.